\documentclass
[prl,print£¬superscriptaddress,showpacs,twocolumn,showkeys]{revtex4}%
\usepackage{amsfonts}
\usepackage{amsmath}
\usepackage{amssymb}
\usepackage{fancyhdr}%
\providecommand{\U}[1]{\protect\rule{.1in}{.1in}}

\begin{document}
\preprint{HEP/123-qed}
\title{\textbf{Extended Bell inequality and maximum violation}}
\author{Yan Gu, Haifeng Zhang, Zhigang Song}
\affiliation{Institute of Theoretical Physics and Department of Physics, State Key
Laboratory of Quantum Optics and Quantum Optics Devices, Shanxi University,
Taiyuan, Shanxi 030006, China}
\author{J. -Q. Liang}
\affiliation{{Institute of Theoretical Physics and Department of Physics, State Key
Laboratory of Quantum Optics and Quantum Optics Devices, Shanxi University,
Taiyuan, Shanxi 030006, China} }
\affiliation{{*jqliang@sxu.edu.cn}}
\author{L. -F. Wei}
\affiliation{State Key Laboratory of Optoelectronic Materials and Technologies, School of
Physics and Engineering, Sun Yat-Sen University, Guangzhou 510275, China}
\affiliation{Quantum Optoelectronics Laboratory, School of Physics and Technology,
Southwest Jiaotong University, Chengdu 610031, China}
\received{17 July 2018}

\revised{2 August 2018}

\begin{abstract}
The original formula of Bell inequality (BI) in terms of two-spin singlet has
to be modified for the entangled-state with parallel spin polarization. Based
on classical statistics of the particle-number correlation, we prove in this
paper an extended BI, which is valid for two-spin entangled states with both
parallel and antiparallel polarizations. The BI and its violation can be
formulated in a unified formalism based on the spin coherent-state quantum
probability statistics with the state-density operator, which is separated to
the local and non-local parts. The local part gives rise to the BI, while the
violation is a direct result of the non-local quantum interference between two
components of entangled state. The Bell measuring outcome correlation denoted
by $P_{B}$ is always less than or at most equal to one for the local realistic
model ($P_{B}^{lc}\leq1$) regardless of the specific superposition
coefficients of entangled state. Including the non-local quantum interference
the maximum violation of BI is found as $P_{B}^{\max}$ $=2$, which, however
depends on state parameters and three measuring directions as well. Our result
is suitable for entangled photon pairs.

\end{abstract}
\maketitle



\thispagestyle{fancy}

\part{{\protect\Large 1. Introduction}}

The non-locality is the most striking characteristic of quantum mechanics
beyond our intuition of space and time in the classical field
theory.\textsuperscript{\cite{1,2,3}} It\textbf{ }has no classical counterpart
and therefore has been receiving continuously theoretical attention ever since
the birth of quantum mechanics.\textbf{ }The two-particle entangled-state as a
typical example of non-locality was originally considered by
Einstein-Podolsky-Rosen to question the complicity of quantum mechanics. It
has become the essential ingredients in quantum information and
computation.\textsuperscript{\cite{4,5,6}} The quantum nonlocal correlation by
local measurements on distant parts of a quantum system is a consequence of
entanglement, which is incompatible with local hidden variable
models.\textsuperscript{\cite{7}} This was discovered by Bell, who further
established a theorem known as Bell inequality (BI)\textsuperscript{\cite{7}}
to provide a possibility of quantitative test for non-local correlations,
which lead necessarily to the violation of BI.\textbf{ }The overwhelming
experimental evidence\textsuperscript{\cite{8,9,10,11,12,13}} for the
violation of BI in some entangled-states invalidates local realistic
interpretations of quantum mechanics. Various extensions of the original BI
have proposed from both theoretical and experimental
viewpoints.\textsuperscript{\cite{14,15,16,17,18,19,20,21}} The nonlocality
has been also justified undoubtedly in various
aspects.{\textsuperscript{\cite{14,15}} Soon after the }pioneer work of Bell,
Clauser-Horne-Shimony-Holt (CHSH) formulated a modified form of the
inequality,\textsuperscript{\cite{22}} which is more suitable for the
quantitative test and therefore attracts most attentions of experiments. An
alternative inequality for the local realistic model\textbf{ }was formulated
by Wigner\textsuperscript{\cite{24,25,26}} known as Wigner inequality (WI),
which needs measurements of particle number probabilities only along one
direction of spin-polarization. It is assumed that the joint probability
distributions for measuring outcomes satisfy the locality
condition\textsuperscript{\cite{27}} in the underlying stochastic hidden
variable space. The experimental evidence strongly supports the quantum
non-locality, however the underlying physical-principle is
obscure.\textsuperscript{\cite{28}} Various aspects relating to the initial
debate remain to be fully understood.

In order to have a better understanding of the underlying physics we in
previous publications{\textsuperscript{\cite{29,30}} formulated} the BI and
its violation in a unified formalism by means of the spin coherent-state
quantum probability statistics along with the assumption of
measurement-outcome-independence. The density operator of a bipartite
entangled-state can be separated into the local and non-local
parts,\textsuperscript{\cite{30}} with which\textbf{ }the measuring outcome
correlation is then evaluated by the quantum probability statistics in the
spin coherent-state base vectors. The local part of density operator gives
rise to the BI, while its violation is a direct result of non-local
correlations of entangled states.\textsuperscript{\cite{30}}\textbf{ }We
predicted a spin parity effect\textsuperscript{\cite{29}} in the violation of
BI, which is violated by the entangled-states of half-integer but not the
integer spins. It was moreover demonstrated that the violation is seen to be
an effect of Berry phase induced by relative-reversal measurements of two spins.

The original formula of BI is actually valid for arbitrary two-spin
entangled-state with antiparallel polarization{\textsuperscript{\cite{31}}}
beyond the singlet state. However, it has to be modified {by the change of a
sign} for the parallel spin polarization.{\textsuperscript{\cite{30}} It is an
interesting question whether or not a unified inequality exists} for both
antiparallel and parallel spin-polarizations. It is the main goal of the
present paper to establish an extended BI valid for two kinds of entangled
states. Following the recent work for the maximum
violation\textsuperscript{\cite{35}} of WI, the maximum violation bound of the
BI is also obtained to demonstrate a fact that the BI with three-direction
measurements is equally convenient as CHSH inequality for the experimental
test. A loophole-free experimental verification of the violation of CHSH
inequality was reported recently by means of electronic spin associated with a
single nitrogen-vacancy defect center in a diamond
chip\textsuperscript{\cite{36,38}} and also for the two-photon entangled
states with mutually perpendicular polarizations\textsuperscript{\cite{37}}.
The formalism and results in the present paper are also suitable for the
entangled photon pairs.

\part{{\protect\Large 2. Spin coherent-state quantum probability statistics
and BI }}

In our\textbf{ }formalism the Bell-type inequalities and their violation are
formulated in a unified manner by means of the spin coherent-state quantum
probability statistics.\textsuperscript{\cite{29,30}} We begin with an
arbitrary two-spin entangled state with antiparallel polarization%
\begin{equation}
|\psi\rangle=c_{1}|+,-\rangle+c_{2}|-,+\rangle, \label{1}%
\end{equation}
in which\textbf{ }$|\pm\rangle$\textbf{ }are considered as the usual
spin-$1/2$ eigenstates $\left(  \hat{\sigma}_{z}|\pm\rangle=\pm|\pm
\rangle\right)  $. The normalized coefficients can be parameterized as\textbf{
}$c_{1}=e^{i\eta}\sin\xi$\textbf{, }$c_{2}=e^{-i\eta}\cos\xi$ in terms of the
arbitrary real parameters $\eta$, $\xi$. \textbf{ }The density operator of an
entangled state can be separated to the local (or classical) and non-local (or
quantum coherent) parts such that\
\begin{equation}
\hat{\rho}=\hat{\rho}_{lc}+\hat{\rho}_{nlc}. \label{2}%
\end{equation}
The local part
\begin{equation}
\hat{\rho}_{lc}=\sin^{2}\xi|+,-\rangle\left\langle +,-\right\vert +\cos^{2}%
\xi|-,+\rangle\left\langle -,+\right\vert , \label{ap}%
\end{equation}
which is the classical two-particle probability-density operator, gives rise
to the local realistic bound of measuring outcome correlation, namely the
BIs.\textbf{ }While the non-local part%
\[
\hat{\rho}_{nlc}=\sin\xi\cos\xi\left(  e^{2i\eta}|+,-\rangle\left\langle
-,+\right\vert +e^{-2i\eta}|-,+\rangle\left\langle +,-\right\vert \right)
\]
describing the quantum coherence between two remote spins results in the
violation of the BIs.\textsuperscript{\cite{29,30}} For the entangled state of
parallel polarization
\[
|\psi\rangle=c_{1}|+,+\rangle+c_{2}|-,-\rangle,
\]
the local and non-local parts of density operator
become\textsuperscript{\cite{30}}
\begin{equation}
\hat{\rho}_{lc}=\sin^{2}\xi|+,+\rangle\left\langle +,+\right\vert +\cos^{2}%
\xi|-,-\rangle\left\langle -,-\right\vert , \label{p}%
\end{equation}
and%
\[
\hat{\rho}_{nlc}=\sin\xi\cos\xi\left(  e^{2i\eta}|+,+\rangle\left\langle
-,-\right\vert +e^{-2i\eta}|-,-\rangle\left\langle +,+\right\vert \right)
\]
respectively.

\emph{2.1 Spin measuring outcome correlation and BI}

We assume to measure two spins independently along two arbitrary directions,
say\ $\mathbf{a}$\ and\ $\mathbf{b}$.\textbf{ }Each measuring outcome falls
necessarily\textbf{ }into the eigenvalues of projection spin-operators
$\hat{\sigma}\cdot\mathbf{a}$\textbf{\ }and\textbf{\ }$\hat{\sigma}%
\cdot\mathbf{b}$\textbf{ }i.e.\textbf{\ }%
\[
\hat{\sigma}\cdot\mathbf{a|}\pm\mathbf{a}\rangle=\pm\mathbf{|}\pm
\mathbf{a}\rangle,\quad\hat{\sigma}\cdot\mathbf{b|}\pm\mathbf{b}\rangle
=\pm\mathbf{|}\pm\mathbf{b}\rangle,
\]
according to the quantum measurement theory.\textbf{ }Solving the above
eigenvalue equations for each direction $\mathbf{r}$ ($\mathbf{r}%
=\mathbf{a},\mathbf{b}$) we obtain two orthogonal eigenstates given by\textbf{
}%
\begin{align}
\left\vert +\mathbf{r}\right\rangle  &  =\cos\frac{\theta_{r}}{2}\left\vert
+\right\rangle +\sin\frac{\theta_{r}}{2}e^{i\phi_{r}}\left\vert -\right\rangle
,\nonumber\\
\left\vert -\mathbf{r}\right\rangle  &  =\sin\frac{\theta_{r}}{2}\left\vert
+\right\rangle -\cos\frac{\theta_{r}}{2}e^{i\phi_{r}}\left\vert -\right\rangle
, \label{3}%
\end{align}
in which\textbf{ }the unit vector $\mathbf{r}=(\sin\theta_{r}\cos\phi_{r}%
,\sin\theta_{r}\sin\phi_{r},\cos\theta_{r})$ is parameterized with the polar
and azimuthal angles\ $\theta_{r},\ \phi_{r}$.\textbf{ }The two orthogonal
states\ $|\pm\mathbf{r}\rangle$\ are known as spin coherent states of north-
and south- pole gauges.\textsuperscript{\cite{32,33,34}}\textbf{ }The
eigenstates of projection spin-operators\textbf{\ }$\hat{\sigma}%
\cdot\mathbf{a}$\textbf{\ }and\textbf{ }$\hat{\sigma}\cdot\mathbf{b}%
$\textbf{\ }form\ a measuring-outcome independent vector base for two spins
measured respectively along the\textbf{\ }$\mathbf{a}$\textbf{, }$\mathbf{b}%
$\textbf{ }directions.\textbf{ }The four base vectors are labeled as%

\begin{equation}
\left\vert 1\right\rangle =\left\vert +\mathbf{a},+\mathbf{b}\right\rangle
,\left\vert 2\right\rangle =\left\vert +\mathbf{a},-\mathbf{b}\right\rangle
,\left\vert 3\right\rangle =\left\vert -\mathbf{a},+\mathbf{b}\right\rangle
,\left\vert 4\right\rangle =\left\vert -\mathbf{a},-\mathbf{b}\right\rangle
\label{vec}%
\end{equation}
for the sake of simplicity.\textbf{ }The measuring outcome
correlation\textsuperscript{\cite{29,30}} is obviously\textbf{ }%
\begin{equation}
P(a,b)=Tr[\hat{\Omega}(a,b)\hat{\rho}]=\rho_{11}-\rho_{22}-\rho_{33}+\rho
_{44}, \label{4}%
\end{equation}
where
\[
\hat{\Omega}(ab)=(\hat{\sigma}\cdot\mathbf{a})(\hat{\sigma}\cdot\mathbf{b}),
\]
is the spin correlation operator and $\rho_{ii}=\langle i|\hat{\rho}|i\rangle$
($i=1,2,3,4$) denote matrix elements of the density operator. The measuring
outcome correlation\textbf{ }can be also separated to local and non-local
parts%
\[
P(a,b)=P_{lc}(a,b)+P_{nlc}(a,b),
\]
with%
\[
P_{lc}(a,b)=Tr[\hat{\Omega}(a,b)\hat{\rho}_{lc}],
\]
and
\[
P_{nlc}(a,b)=Tr[\hat{\Omega}(a,b)\hat{\rho}_{nlc}].
\]
Submitting the local parts of density operators Eq.(\ref{ap}), Eq.(\ref{p})
into the local measuring-outcome correlation $P_{lc}(a,b)$ we have
\[
P_{lc}(a,b)=\mp\cos\theta_{a}\cos\theta_{b},
\]
respectively for the antiparallel and parallel spin polarizations. The BI
becomes correspondingly\textsuperscript{\cite{29,30}}
\[
1\pm P_{lc}(b,c)\geq\left\vert P_{lc}(a,b)-P_{lc}(a,c)\right\vert
\]
for the antiparallel and parallel entangled states.

\emph{2.2 Particle-number correlation probability}

In the Wigner formalism\textsuperscript{\cite{35}} the particle-number
correlation probability is considered instead of the spin measuring-outcome
correlation.\textbf{ }The quantity defined by%
\begin{equation}
N(+a,+b)=|\langle+a,+b|\psi\rangle|^{2}=\langle+a,+b|\hat{\rho}|+a,+b\rangle
=\rho_{11} \label{N}%
\end{equation}
describes the particle-number correlation probability for two positive-spin
particles measured respectively along\textbf{ }$\mathbf{a}$,\textbf{
}$\mathbf{b}$\textbf{ }directions. Correspondingly we have
\begin{equation}
N(+a,-b)=\rho_{22},\quad N(-a,+b)=\rho_{33},\quad N(-a,-b)=\rho_{44},
\label{N1}%
\end{equation}
which are all positive quantities different from the spin measuring-outcome
correlations.\textbf{ }The spin measuring-outcome correlation $P(a,b)$ in the
BI are related to the four particle-number correlation probabilities by%
\begin{equation}
P(a,b)=N(+a,+b)-N(+a,-b)-N(-a,+b)+N(-a,-b), \label{smc}%
\end{equation}
which is the difference between the particle number probabilities of same
direction measurement and that of opposite directions.

\part{{\protect\Large 3. Extended BI and maximum violation }}

The extended BI for both parallel and antiparallel polarizations is obviously%
\[
1+\left\vert P_{lc}(b,c)\right\vert \geq\left\vert P_{lc}(a,b)-P_{lc}%
(a,c)\right\vert ,
\]
for the local model, since
\[
1+\left\vert P_{lc}(b,c)\right\vert \geq1\pm P_{lc}(b,c).
\]
We define a quantum Bell correlation probability (QBCP) that%
\[
P_{B}=\left\vert P(a,b)-P(a,c)\right\vert -\left\vert P(b,c)\right\vert .
\]
The extended BI is then%
\[
P_{B}^{lc}\leq1,
\]
which is violated once $P_{B}>1$.

In the Appendix we specifically present a simple proof of the validity of the
extended BI, in terms of the classical statistics with the particle-number
correlation probabilities in the Wigner
formulation.\textsuperscript{\cite{24,25,26}} A interesting question is to
find the maximum violation bound, which is useful for the experimental verification.

\emph{3.1 Two-spin entangled state with antiparallel polarization}

By means of the spin coherent-state quantum probability statistics we can
obtain quantum correlation probability $P\left(  a,b\right)  $ for the
two-spin entangled state with antiparallel polarization. The entire (quantum)
correlation-probability including the non-local parts becomes%
\[
P(a,b)=-\cos\theta_{a}\cos\theta_{b}+\sin2\xi\sin\theta_{a}\sin\theta_{b}%
\cos\left(  \phi_{a}-\phi_{b}+2\eta\right)  .
\]
The QBCP for the three-direction measurement is found as%
\begin{align}
P_{B}  &  =|-\cos\theta_{a}\cos\theta_{b}+\sin2\xi\sin\theta_{a}\sin\theta
_{b}\cos\left(  \phi_{a}-\phi_{b}+2\eta\right) \nonumber\\
&  +\cos\theta_{a}\cos\theta_{c}-\sin2\xi\sin\theta_{a}\sin\theta_{c}%
\cos\left(  \phi_{a}-\phi_{c}+2\eta\right)  |\nonumber\\
&  -\left\vert -\cos\theta_{b}\cos\theta_{c}+\sin2\xi\sin\theta_{b}\sin
\theta_{c}\cos\left(  \phi_{b}-\phi_{c}+2\eta\right)  \right\vert .
\label{QBCP}%
\end{align}
Since the polar angle $\theta$ is restricted between $0$ and $\pi$, the
quantity $\sin\theta_{a}\sin\theta_{b}$ is great than or equal to zero. We
then obtain after a simple algebra the inequality of QBCP%
\begin{equation}
P_{B}\leq|-\cos\left(  \theta_{a}\pm\theta_{b}\right)  +\cos\left(  \theta
_{a}\mp\theta_{c}\right)  |. \label{ine}%
\end{equation}
Thus we have the maximum violation bound
\[
P_{B}^{\max}=2.
\]
\bigskip As a matter of fact the QBCP is bounded by\textbf{ }$2\geq P_{B}%
\geq-1$.

For the measuring directions with polar and azimuthal angles $\theta
_{a}=\theta_{b}=\theta_{c}=\pi/2$ and $\phi_{a}=\pi/2,\phi_{b}=0,\phi_{c}%
=\pi,$ the QBCP Eq.(\ref{QBCP}) becomes%
\[
P_{B}=2\left\vert \sin2\xi\sin\left(  2\eta\right)  \right\vert -\left\vert
\sin2\xi\cos\left(  2\eta\right)  \right\vert .
\]
The three directions of measurements $\mathbf{a}$, $\mathbf{b}$, $\mathbf{c}$
are set up with $\mathbf{a}$ along positive $y$-axis, $\mathbf{b}$,
$\mathbf{c}$ along positive and negative $x$-axis respectively. The maximum
violation $P_{B}^{\max}=2$ is approached with the state parameters, for
example, $\xi=(\pi/4)\operatorname{mod}2\pi,\eta=(\pi/4)\operatorname{mod}%
2\pi$. The entangled state in this case is
\[
|\psi\rangle=\frac{1}{\sqrt{2}}\left(  e^{i\frac{\pi}{4}}|+,-\rangle
+e^{-i\frac{\pi}{4}}|-,+\rangle\right)  .
\]
The violation value depends not only on the entangled-state parameters $\xi$,
$\eta$ in our parametrization but also on the three directions of measurements.

Particularly for the two-spin singlet state\textbf{ }%
\[
|\psi_{s}\rangle=\frac{1}{\sqrt{2}}(|+,-\rangle-|-,+\rangle)
\]
with\textbf{ }the state parameters $\xi=3\pi/4,\eta=0$ the QBCP value is found
from Eq.(\ref{QBCP}) as%
\[
P_{B}=|-\cos\left(  \phi_{a}-\phi_{b}\right)  +\cos\left(  \phi_{a}-\phi
_{c}\right)  |-\left\vert \cos\left(  \phi_{b}-\phi_{c}\right)  \right\vert
\]
for the polar angles of three-direction measurements $\theta_{a}=\theta
_{b}=\theta_{c}=\pi/2.$ The maximum QBCP value of two-spin singlet is obtained
as
\[
P_{B}=\sqrt{2},
\]
for azimuthal angles $\phi_{a}=3\pi/4,\phi_{b}=\pi/2,\phi_{c}=0$. It is less
than the maximum violation bound $P_{B}^{\max}=2$ different from the common
believe that the spin-singlet gives rise to the maximum violation bound. To
obtain the violation value $\sqrt{2}$ for the spin singlet state the vector
$\mathbf{b}$ is perpendicular to $\mathbf{c}$ and the vector $\mathbf{a}$ is
parallel to the vector difference $(\mathbf{b-c)}$.

\emph{3.2 Parallel polarization}

For the two-spin entangled state with parallel
polarization{\textsuperscript{\cite{30}}}%
\[
\left\vert \psi\right\rangle =c_{1}\left\vert +,+\right\rangle +c_{2}%
\left\vert -,-\right\rangle
\]
the entire correlation-probability including the non-local parts becomes%
\[
P(a,b)=\cos\theta_{a}\cos\theta_{b}+\sin2\xi\sin\theta_{a}\sin\theta_{b}%
\cos\left(  \phi_{a}+\phi_{b}+2\eta\right)  .
\]
The QBCP then is%
\begin{align}
P_{B}  &  =|\cos\theta_{a}\cos\theta_{b}+\sin2\xi\sin\theta_{a}\sin\theta
_{b}\cos\left(  \phi_{a}+\phi_{b}+2\eta\right) \nonumber\\
&  -\cos\theta_{a}\cos\theta_{c}-\sin2\xi\sin\theta_{a}\sin\theta_{c}%
\cos\left(  \phi_{a}+\phi_{c}+2\eta\right)  |\nonumber\\
&  -\left\vert \cos\theta_{b}\cos\theta_{c}+\sin2\xi\sin\theta_{b}\sin
\theta_{c}\cos\left(  \phi_{b}+\phi_{c}+2\eta\right)  \right\vert ,
\label{QBCP1}%
\end{align}
from which we have the same inequality of QBCP as Eq.(\ref{ine}). Thus, the
maximum value of QBCP is still $P_{B}^{\max}=2.$

The maximum violation of BI can be realized from Eq.(\ref{QBCP1}) for the
parallel polarization state given by%
\[
|\psi\rangle=\frac{1}{\sqrt{2}}\left(  e^{i\frac{\pi}{4}}|+,+\rangle
+e^{-i\frac{\pi}{4}}|-,-\rangle\right)
\]
with the state parameters $\xi=(\pi/4)\operatorname{mod}2\pi,\eta
=(\pi/4)\operatorname{mod}2\pi$. The three-direction measurements should be
arranged respectively with the polar and azimuthal angles $\theta_{a}%
=\theta_{b}=\theta_{c}=\pi/2,$ $\phi_{a}=\pi/2,\phi_{b}=0,\phi_{c}=\pi$.
Namely $\mathbf{a},\mathbf{b},\mathbf{c}$ are perpendicular to the original
spin polarization ($z$-axis) with $\mathbf{a}$ along $y$-directions;
$\mathbf{b}$, $\mathbf{c}$ along $\pm x$-directions.

In conclusion the value of QBCP is restricted by $-1\leq P_{B}\leq2$ for
two-spin entangled states with both parallel and antiparallel polarizations.
The extended BI is violated if $1<P_{B}$. \

\part{{\protect\Large 4. Polarization-entangled photon pairs}}

Polarization-entangled photon pairs play an important role in various quantum
information experiments instead of the two-spin entangled state. We
reformulate the extended BI and maximum violation in terms of our formalism.

\emph{4.1 Entangled photon pairs with mutually perpendicular polarizations}

Two perpendicular polarization states of a single photon may be denoted by
$|e_{x}\rangle$ and $|e_{y}\rangle$ in our framework, where we have assumed
that the polarization plane is perpendicular to the $z$-axis. The entangled
state of a photon pair with mutually perpendicular polarizations can be
represented as
\[
|\psi\rangle=c_{1}|e_{x},e_{y}\rangle+c_{2}|e_{y},e_{x}\rangle,
\]
which corresponds to the two-spin entangled state with antiparallel
spin-polarizations. The normalized coefficients $c_{1}$,\textbf{ }$c_{2}$ are
parameterized as in the spin case. The local and non-local parts of density
operator are the same, however, with the spin states $|\pm\rangle$ replaced
respectively by the photon polarization states $|e_{x}\rangle$, $|e_{y}%
\rangle$. The entangled photon pairs are measured in three arbitrary
directions, say $\mathbf{a},\mathbf{b}\ $and $\mathbf{c}$, in the plane also
perpendicular to the $z$-axis. With respect to a measuring direction, say
$\mathbf{r}=(\cos\phi_{r},\sin\phi_{r},0)$ ($\mathbf{r}=\mathbf{a}%
,\mathbf{b},\mathbf{c}$), the horizontal ($h$) and vertical ($v$) polarization
states are represented as%
\begin{align*}
\left\vert \mathbf{r}_{h}\right\rangle  &  =\cos\phi_{r}\left\vert
e_{x}\right\rangle +\sin\phi_{r}\left\vert e_{y}\right\rangle ,\\
\left\vert \mathbf{r}_{v}\right\rangle  &  =-\sin\phi_{r}\left\vert
e_{x}\right\rangle +\cos\phi_{r}\left\vert e_{y}\right\rangle ,
\end{align*}
where $\phi_{r}$ is the azimuthal angle of the measuring direction
$\mathbf{r}$. In the measuring-outcome independent vector base denoted
similarly by%
\[
\left\vert 1\right\rangle =\left\vert \mathbf{a}_{h},\mathbf{b}_{h}%
\right\rangle ,\left\vert 2\right\rangle =\left\vert \mathbf{a}_{h}%
,\mathbf{b}_{v}\right\rangle ,\left\vert 3\right\rangle =\left\vert
\mathbf{a}_{v},\mathbf{b}_{h}\right\rangle ,\left\vert 4\right\rangle
=\left\vert \mathbf{a}_{v},\mathbf{b}_{v}\right\rangle ,
\]
the local part of the density-operator elements becomes%
\[
\rho_{11}^{lc}=\sin^{2}\xi\cos^{2}\phi_{a}\sin^{2}\phi_{b}+\cos^{2}\xi\sin
^{2}\phi_{a}\cos^{2}\phi_{b},
\]%
\[
\rho_{22}^{lc}=\sin^{2}\xi\cos^{2}\phi_{a}\cos^{2}\phi_{b}+\cos^{2}\xi\sin
^{2}\phi_{a}\sin^{2}\phi_{b},
\]%
\[
\rho_{33}^{lc}=\sin^{2}\xi\sin^{2}\phi_{a}\sin^{2}\phi_{b}+\cos^{2}\xi\cos
^{2}\phi_{a}\cos^{2}\phi_{b},
\]%
\[
\rho_{44}^{lc}=\sin^{2}\xi\sin^{2}\phi_{a}\cos^{2}\phi_{b}+\cos^{2}\xi\cos
^{2}\phi_{a}\sin^{2}\phi_{b}.
\]
The non-local part is%

\[
\rho_{11}^{nlc}=\rho_{44}^{nlc}=-\rho_{22}^{nlc}=-\rho_{33}^{nlc}=\frac{1}%
{4}\sin2\xi\cos2\eta\sin2\phi_{a}\sin2\phi_{b}.
\]
The local measuring-outcome correlation is%

\begin{equation}
P_{lc}(a,b)=\rho_{11}^{lc}-\rho_{22}^{lc}-\rho_{33}^{lc}+\rho_{44}^{lc}%
=-\cos2\phi_{a}\cos2\phi_{b}, \label{perpan}%
\end{equation}
with which it is easy to verify the extended BI%
\begin{equation}
P_{B}^{lc}\leq\left\vert \cos2\phi_{b}-\cos2\phi_{c}\right\vert -\left\vert
\cos2\phi_{b}\cos2\phi_{c}\right\vert \leq1. \label{local}%
\end{equation}
Including the nonlocal part%

\[
P_{nlc}(a,b)=\sin2\xi\sin2\phi_{a}\sin2\phi_{b}\cos2\eta,
\]
the entire correlation-probability becomes%
\[
P(a,b)=-\cos2\phi_{a}\cos2\phi_{b}+\sin2\xi\cos2\eta\sin2\phi_{a}\sin2\phi
_{b}.
\]
The QBCP is
\begin{align}
P_{B}  &  =|-\cos2\phi_{a}\cos2\phi_{b}+\sin2\xi\cos2\eta\sin2\phi_{a}%
\sin2\phi_{b}\nonumber\\
&  +\cos2\phi_{a}\cos2\phi_{c}-\sin2\xi\cos2\eta\sin2\phi_{a}\sin2\phi
_{c}|\nonumber\\
&  -\left\vert -\cos2\phi_{b}\cos2\phi_{c}+\sin2\xi\cos2\eta\sin2\phi_{b}%
\sin2\phi_{c}\right\vert , \label{QBCP2}%
\end{align}
which is then bounded by%
\begin{align*}
P_{B}  &  \leq\left\vert -\cos\left(  2\phi_{a}+2\phi_{b}\right)  +\cos\left(
2\phi_{a}+2\phi_{c}\right)  \right\vert \\
&  \leq P_{B}^{\max}=2.
\end{align*}
From the general form of QBCP Eq.(\ref{QBCP2}) it is easy to check that the
entangled state
\begin{equation}
|\psi\rangle=\frac{1}{\sqrt{2}}\left(  |e_{x},e_{y}\rangle+|e_{y},e_{x}%
\rangle\right)  \label{trip}%
\end{equation}
results in the maximum violation bound $P_{B}^{\max}=2$ for the
three-direction measurements with $\phi_{a}=\pi/8,\phi_{b}=3\pi/8,$ and
$\phi_{c}=15\pi/8$. Namely, $\mathbf{b}$ is perpendicular to $\mathbf{c}$, and
the angle between $\mathbf{a}$ and $\mathbf{c}$ equals $\pi/4$. It is
remarkably to find that the state
\[
\mathbf{\ }|\psi_{s}\rangle=\frac{1}{\sqrt{2}}(|e_{x},e_{y}\rangle
-|e_{y},e_{x}\rangle),
\]
which may be regarded as the the counterpart of two-spin singlet, gives rise
to the QBCP-value again $\sqrt{2}$ less than the maximum bound $P_{B}^{\max}$.
The three angles of measuring directions should be arranged as $\phi_{a}%
=3\pi/8,\phi_{b}=\pi/4,$ and $\phi_{c}=0$ in this state.

\emph{4.2 Parallel polarization }

The state of entangled photon pairs with mutually parallel polarizations is%
\[
|\psi\rangle=c_{1}|e_{x},e_{x}\rangle+c_{2}|e_{y},e_{y}\rangle.
\]
With the same calculation procedure we obtain the measuring outcome
correlation%
\[
P_{lc}(a,b)=\cos2\phi_{a}\cos2\phi_{b},
\]
which has a sign difference with the perpendicular case of Eq.(\ref{perpan}).
The QBCP is invariant as Eq.(\ref{local}) comparing with the entangled state
with perpendicular polarizations, so is the extended BI. Including the
nonlocal part $P_{nlc}(a,b)=\sin2\xi\sin2\phi_{a}\sin2\phi_{b}\cos2\eta$ the
entire correlation-probability is%
\[
P(a,b)=\cos2\phi_{a}\cos2\phi_{b}+\sin2\xi\sin2\phi_{a}\sin2\phi_{b}\cos
2\eta.
\]
The QBCP then is%
\begin{subequations}
\begin{align}
P_{B}  &  =|\cos2\phi_{a}\cos2\phi_{b}+\sin2\xi\sin2\phi_{a}\sin2\phi_{b}%
\cos2\eta\nonumber\\
&  -\cos2\phi_{a}\cos2\phi_{c}-\sin2\xi\sin2\phi_{a}\sin2\phi_{c}\cos
2\eta|\nonumber\\
&  -\left\vert \cos2\phi_{b}\cos2\phi_{c}+\sin2\xi\sin2\phi_{b}\sin2\phi
_{c}\cos2\eta\right\vert , \label{QBCP3}%
\end{align}
which leads again to the maximum violation bound $P_{B}\leq P_{B}^{\max}=2.$

In the entangled state of equal polarizations
\end{subequations}
\[
|\psi\rangle=\frac{1}{\sqrt{2}}\left(  |e_{x},e_{x}\rangle-|e_{y},e_{y}%
\rangle\right)  ,
\]
the maximum violation $P_{B}^{\max}=2$ can be achieved from Eq.(\ref{QBCP3})
for the same measuring directions of $\mathbf{a}$, $\mathbf{b}$, and
$\mathbf{c}$ as in the state of Eq.(\ref{trip}).

\part{{\protect\Large 5. Conclusions and discussions}}

The BI and its violation are formulated in a unified way by the spin
coherent-state quantum probability statistics, in which the state density
operator is separated to the local and nonlocal parts. The BI is a direct
result of local model, while the nonlocal part from the coherent interference
of two components of the entangled state leads to the violation. The original
BI, which was derived from the two-spin singlet, is extended to a unified form
valid for the general entangled state with both antiparallel and parallel
polarizations. Up to date the experimental
test\textsuperscript{\cite{36,38,37}} of the inequality violation are mainly
focused on the CHSH form, which provides a qualitative bound of the violation.
The maximum violation value $P_{B}^{\max}$ $=2$ of the extended BI is two
times of BI bound unit one ($P_{B}^{lc}\leq1$), while it is $\sqrt{2}$ times
in the CHSH inequality case. We thus conclude that the extended BI is at least
equally convenient for the experimental verification of its violation, which
is expected in the future experiments. We moreover demonstrate that the
violation value depends not only on the measuring directions but also two
superposition coefficients of the entangled states, namely the angle
parameters $\xi$, $\eta$ in our formalism. It is remarkably to find that the
maximum violation for the spin singlet is only $\sqrt{2}$ less than the
maximum violation bound $P_{B}^{\max}$. Our observation is different from the
common believe that the spin-singlet would give rise to the maximum violation.
The extended BI and violation are also suitable to entangled photon pairs. The
BI and WI were formulated respectively with the spin and
particle-number-probability correlations in the literature. The two measuring
outcome correlations are also unified in our formalism of quantum probability statistics.\textsuperscript{\cite{29,30,35}}

\part{{\protect\large Acknowledge}}

This work was supported in part by National Natural Science Foundation of
China, under Grants No. 11275118, U1330201.

\part{{\protect\large Appendix}}

Extended BI from viewpoint of classical statistics with realistic model by
means of the measuring outcome correlation of particle number probabilities.

\emph{A1. Two-spin entangled state with antiparallel polarization}

In this case there are eight independent (measuring-outcome) particle-number
probabilities given by the following table.\textsuperscript{\cite{25}}

\ {\footnotesize
\[
Table\ A1.\ Antiparallel
\]
}%
\[
{\footnotesize \ }{\footnotesize
\begin{tabular}
[c]{ccc}\hline
population & particle1 & particle2\\\hline
$N_{1}$ & $\left(  +a,+b,+c\right)  $ & $\left(  -a,-b,-c\right)  $\\
$N_{2}$ & $\left(  +a,+b,-c\right)  $ & $\left(  -a,-b,+c\right)  $\\
$N_{3}$ & $\left(  +a,-b,+c\right)  $ & $\left(  -a,+b,-c\right)  $\\
$N_{4}$ & $\left(  +a,-b,-c\right)  $ & $\left(  -a,+b,+c\right)  $\\
$N_{5}$ & $\left(  -a,+b,+c\right)  $ & $\left(  +a,-b,-c\right)  $\\
$N_{6}$ & $\left(  -a,+b,-c\right)  $ & $\left(  +a,-b,+c\right)  $\\
$N_{7}$ & $\left(  -a,-b,+c\right)  $ & $\left(  +a,+b,-c\right)  $\\
$N_{8}$ & $\left(  -a,-b,-c\right)  $ & $\left(  +a,+b,+c\right)  $\\\hline
\end{tabular}
\ }%
\]
${\footnotesize \ \ \ }\ ${\footnotesize \vspace{2mm} }Table A1 lists for the
measuring outcomes of two spins respectively along three directions\textbf{
}$\mathbf{a}$\textbf{, }$\mathbf{b}$\textbf{ }and\textbf{ }$\mathbf{c}%
$\textbf{. }For the local realistic model the measuring outcome correlation
probabilities can be expressed in terms of the above population probabilities
such that

$\ \ \ \ \ \ \ \ \ \ \ \ \ \ \ \ \ \ \ \ \ $%
\[
N_{lc}\left(  +a,+b\right)  =\frac{\left(  N_{3}+N_{4}\right)  }{\sum_{i}%
^{8}N_{i}},
\]%
\[
N_{lc}\left(  -a,-b\right)  =\frac{\left(  N_{5}+N_{6}\right)  }{\sum_{i}%
^{8}N_{i}},
\]%
\[
N_{lc}\left(  +a,-b\right)  =\frac{\left(  N_{1}+N_{2}\right)  }{\sum_{i}%
^{8}N_{i}},
\]
and%
\[
N_{lc}\left(  -a,+b\right)  =\frac{\left(  N_{7}+N_{8}\right)  }{\sum_{i}%
^{8}N_{i}}.
\]
The spin measuring outcome correlation Eq.(\ref{smc}) in BI is related to the
four particle-number probabilities as%
\begin{align*}
P_{lc}\left(  a,b\right)   &  =N_{lc}\left(  +a,+b\right)  +N_{lc}\left(
-a,-b\right)  -N_{lc}\left(  +a,-b\right) \\
&  -N_{lc}\left(  -a,+b\right) \\
&  =\frac{1}{\sum_{i}^{8}N_{i}}\left(  N_{3}+N_{4}+N_{5}+N_{6}-N_{1}%
-N_{2}-N_{7}-N_{8}\right)  .
\end{align*}
Correspondingly for particle-$1$ measured along $a$, particle-$2$ along $c$
and particle-$1$ along $b$, particle-$2$ along $c$ the measuring outcome
correlations are similarly obtained as
\begin{align*}
P_{lc}\left(  a,c\right)   &  =N_{lc}\left(  +a,+c\right)  +N_{lc}\left(
-a,-c\right)  -N_{lc}\left(  +a,-c\right) \\
&  -N_{lc}\left(  -a,+c\right) \\
&  =\frac{1}{\sum_{i}^{8}N_{i}}\left(  N_{2}+N_{4}+N_{5}+N_{7}-N_{1}%
-N_{3}-N_{6}-N_{8}\right)  .
\end{align*}
and%
\begin{align*}
P_{lc}\left(  b,c\right)   &  =N_{lc}\left(  +b,+c\right)  +N_{lc}\left(
-b,-c\right)  -N_{lc}\left(  +b,-c\right) \\
&  -N_{lc}\left(  -b,+c\right) \\
&  =\frac{1}{\sum_{i}^{8}N_{i}}\left(  N_{2}+N_{6}+N_{3}+N_{7}-N_{1}%
-N_{5}-N_{4}-N_{8}\right)  .
\end{align*}
We then have%
\[
P_{lc}\left(  a,b\right)  -P_{lc}\left(  a,c\right)  =\frac{2}{\sum_{i}%
^{8}N_{i}}\left[  N_{3}+N_{6}-N_{2}-N_{7}\right]  .
\]
Thus Bell correlation becomes%
\begin{align}
P_{B}^{lc}  &  =\left\vert P_{lc}(a,b)-P_{lc}(a,c)\right\vert -\left\vert
P_{lc}(b,c)\right\vert \nonumber\\
&  =\frac{1}{\sum_{i}^{8}N_{i}}[2\left\vert N_{3}+N_{6}-N_{2}-N_{7}\right\vert
\nonumber\\
&  -\left\vert N_{2}+N_{6}+N_{3}+N_{7}-N_{1}-N_{5}-N_{4}-N_{8}\right\vert ].
\end{align}
It is easy to verify the inequality
\[
P_{B}^{lc}\leq1.
\]
The equality holds only in the special cases when $N_{2}=N_{7}=0$ or
$N_{3}=N_{6}=0$.

\emph{A2. Two-spin entangled state with parallel polarization}

For the entangled state with parallel spin-polarization the eight independent
particle-number probabilities become {\textsuperscript{\cite{35}} }those
listed in the following table.%
\[
{\footnotesize Table\ A2.\ Spin\ correlation\ Measurements}%
\]%
\[
{\footnotesize
\begin{tabular}
[c]{ccc}\hline
population & particle1 & particle2\\\hline
$N_{1}$ & $\left(  +a,+b,+c\right)  $ & $\left(  +a,+b,+c\right)  $\\
$N_{2}$ & $\left(  +a,+b,-c\right)  $ & $\left(  +a,+b,-c\right)  $\\
$N_{3}$ & $\left(  +a,-b,+c\right)  $ & $\left(  +a,-b,+c\right)  $\\
$N_{4}$ & $\left(  +a,-b,-c\right)  $ & $\left(  +a,-b,-c\right)  $\\
$N_{5}$ & $\left(  -a,+b,+c\right)  $ & $\left(  -a,+b,+c\right)  $\\
$N_{6}$ & $\left(  -a,+b,-c\right)  $ & $\left(  -a,+b,-c\right)  $\\
$N_{7}$ & $\left(  -a,-b,+c\right)  $ & $\left(  -a,-b,+c\right)  $\\
$N_{8}$ & $\left(  -a,-b,-c\right)  $ & $\left(  -a,-b,-c\right)  $\\\hline
\end{tabular}
\ \ }%
\]
${\footnotesize \ \ }$The measuring outcome correlation probabilities then
are\ \ \ \ \ \ \ \ \ \ \ \ \ \ \ \ \ \ \ \ \ \ \ \ \ \ \ \ \ \ \ \ $\ \ \ \ \ \ \ \ \ \ \ \ \ \ \ \ \ \ \ \ \ \ \ \ \ \ \ \ \ \ \ \ \ \ \ \ \ \ $%
\[
N_{lc}\left(  +a,+b\right)  =\frac{\left(  N_{1}+N_{2}\right)  }{\sum_{i}%
^{8}N_{i}},
\]%
\[
N_{lc}\left(  -a,-b\right)  =\frac{\left(  N_{7}+N_{8}\right)  }{\sum_{i}%
^{8}N_{i}},
\]%
\[
N_{lc}\left(  +a,-b\right)  =\frac{\left(  N_{3}+N_{4}\right)  }{\sum_{i}%
^{8}N_{i}},
\]
and%
\[
N_{lc}\left(  -a,+b\right)  =\frac{\left(  N_{5}+N_{6}\right)  }{\sum_{i}%
^{8}N_{i}}.
\]
Repeat the same calculation procedure as in the antiparallel case we again
have the extended BI $P_{B}^{lc}\leq1$.

\part{{\protect\large References}}


\begin{thebibliography}{99}                                                                                               %


\bibitem {1}{\small Su H Y, Wu Y C and Chen J L 2013 \emph{Phys. Rev. A}
\textbf{88} 022124.}

\bibitem {2}{\small Kwiat P G, Barraza-Lopez S, Stefanov A, \emph{et al.} 2001
\emph{Nature} \textbf{409}(6823) 1014-1017.}

\bibitem {3}{\small Popescu S 2010 \emph{Nat. Phys.} \textbf{6}(3) 151-153. }

\bibitem {4}{\small Nielsen M A and Chuang I L 2011 \textit{Quantum
Computation and Quantum Information} (Cambridge: Cambridge University Press)
pp. 1--59. }

\bibitem {5}{\small Bennett C H and Divincenzo D P 2000 \emph{Nature}
\textbf{404}(6775) 247.}

\bibitem {6}{\small Loss D and DiVincenzo D P 1998 \emph{Phys. Rev. A}
\textbf{57}(1) 120.}

\bibitem {7}{\small Bell J S 1964 \textit{Physics} \textbf{1} 195.}

\bibitem {8}{\small Waldherr G, \emph{et al}. 2011 \emph{Phys. Rev. Lett.}
\textbf{107} 090401.}

\bibitem {9}{\small Sakai H, \emph{et al.} 2006 \emph{Phys. Rev. Lett.}
\textbf{97} 150405.}

\bibitem {10}{\small Pal K F and Vertesi T 2017 \emph{Phys. Rev. A}
\textbf{96} 022123.}

\bibitem {11}{\small Rowe M A, Kielpinski D, Meyer V, \emph{et al}. 2001
\emph{Nature} \textbf{409}(6822) 791-794.}

\bibitem {12}{\small Dada A C, Leach J, Buller G S, \emph{et al}. 2011
\emph{Nat. Phys.} \textbf{7}(9) 677-680.}

\bibitem {13}{\small Gisin N and Peres A 1992 \emph{Phys. Lett. A}
\textbf{162}(1) 15-17.}

\bibitem {14}{\small Brito S G A, Amaral B and Chaves R 2018 \emph{Phys. Rev.
A} \textbf{97} 022111.}

\bibitem {15}{\small Pozsgay V, Hirsch F, Branciard C, \emph{et al}. 2017
\emph{Phys. Rev. A }\textbf{96} 062128.}

\bibitem {16}{\small Groblacher S, Paterek T, Kaltenbaek R, \emph{et al}. 2007
\emph{Nature} \textbf{446} 871.}

\bibitem {17}{\small Buhrman H, Cleve R, Massar S, \emph{et al}. 2010
\emph{Rev. Mod. Phys.} \textbf{82}(1) 665.}

\bibitem {18}{\small Cabello A and Sciarrino F 2012 \emph{Phys. Rev. X}
\textbf{2} 021010.}

\bibitem {19}{\small Wei L F, Liu Y X, and Nori F 2005 \emph{Phys. Rev. B}
\textbf{72} 104516.}

\bibitem {20}{\small Garcia-Patron R, Fiurasek J, Cerf N J, \emph{et al}. 2004
\emph{Phys. Rev. Lett.} \textbf{93}(13) 130409.}

\bibitem {21}{\small Hess K, Raedt H D and Michielsen K 2017 \emph{J. Mod.
Phys.} \textbf{8} 57.}

\bibitem {22}{\small Clauser J F, Horne M A, Shimony A and Holt R A 1969
\emph{Phys. Rev. Lett.} \textbf{23} 880.}

\bibitem {24}{\small Wigner E P 1970 \emph{Am. J. Phys.}\textbf{ 38}(8)
1005-1009.}

\bibitem {25}{\small Sakurai J J, Tuan S F and Commins E D 1995 \emph{Am. J.
Phys.} \textbf{63}(63) 93-95.}

\bibitem {26}{\small Home D, Saha D and Das S\ 2015 \emph{Phys. Rev. A}
\textbf{91}(1) 012102.}

\bibitem {27}{\small Das D, Datta S, Goswami S, Majumdar A S and Home D\ 2017
\emph{Phys. Lett. A }\textbf{381}(39) 3396-3404.}

\bibitem {28}{\small Pawlowski M, Paterek T, Kaszlikowski D, Scarani
V,\ Winter A and Zukowski M 2009 \emph{Nature} \textbf{461} 1101.}

\bibitem {29}{\small Song Z, Liang J-Q and Wei L-F 2014 \emph{Mod. Phys. Lett.
B} \textbf{28}(01) 1450004.}

\bibitem {30}{\small Zhang H, Wang J, Song Z, \emph{et al}. 2017 \emph{Mod.
Phys. Lett. B} \textbf{31}(04) 1750032.}

\bibitem {31}{\small Grimme S 2003\emph{ J. Chem. Phys.} \textbf{118}(20)
9095-9102.}

\bibitem {35}{\small Gu Y, Zhang H, Song Z, Liang J-Q and Wei L-F 2018
\textit{Int. J. Quantum Inf.} \textbf{16} 1850041.}

\bibitem {36}{\small Hensen B, Bernien H, Dr\.{e}au A E, \textit{et al}. 2015
\textit{Nature} \textbf{526} 682.}

\bibitem {38}{\small Hensen B, Kalb N, Blok M S, \textit{et al}. 2016
\textit{Sci. Rep.} \textbf{6} 30289.}

\bibitem {37}{\small Yin J, Cao Y, Li Y H, \textit{et al.} 2017
\textit{Science} \textbf{356} 1140--1144.}

\bibitem {32}{\small Liang J Q and Wei L F 2011 \textit{New Advances in
Quantum Physics} (Beijing: Science Press) p. 56.}

\bibitem {33}{\small Bai X-M, Gao C-P, Li J-Q and Liang J-Q 2017 \emph{Opt.
Express} \textbf{25} 17051-17065.}

\bibitem {34}{\small Zhao X-Q, Liu N and Liang J-Q 2014 \emph{Phys. Rev. A}
\textbf{90} 023622.}
\end{thebibliography}
\end{document}